# Directivity modulation of exciton emission using single dielectric nanospheres

Jie Fang[1,*], Mingsong Wang[1,2,*,✉], Kan Yao[1,*], Tianyi Zhang[3], Alex Krasnok[2], Taizhi Jiang[4], Junho Choi[5], Ethan Kahn[3], Brian A. Korgel[4], Mauricio Terrones[3], Xiaoqin Li[5], Andrea Alù[2,✉], and Yuebing Zheng[1,✉]

[1]Walker Department of Mechanical Engineering and Texas Materials Institute, The University of Texas at Austin, Austin, TX 78712, USA

[2]Photonics Initiative, Advanced Science Research Center, City University of New York, New York, NY 10031, USA

[3]Department of Materials Science and Engineering, Department of Physics, Department of Chemistry and Center for 2-Dimensional and Layered Materials, The Pennsylvania State University, University Park, PA 16802, USA

[4]McKetta Department of Chemical Engineering, The University of Texas at Austin, Austin, TX 78712, USA.

[5]Department of Physics, The University of Texas at Austin, Austin, TX 78712, USA

*These authors contributed equally: Jie Fang, Mingsong Wang, Kan Yao

✉Email: mwang2@gc.cuny.edu, aalu@gc.cuny.edu, zheng@austin.utexas.edu

**Abstract** Coupling emitters with nanoresonators is an effective strategy to control light emission at the subwavelength scale with high efficiency. Low-loss dielectric nanoantennas hold particular promise for this purpose, owing to their strong Mie resonances. Herein, we explore a highly miniaturized platform for the control of emission based on individual subwavelength Si nanospheres (SiNSs) to modulate the directional excitation and exciton emission of two-dimensional transition metal dichalcogenides (2D TMDs). A modified Mie theory for dipole-sphere hybrid systems is derived to instruct the optimal design for desirable modulation performance. Controllable forward-to-backward intensity ratios are experimentally validated in 532 nm laser excitation and 635 nm exciton emission from a monolayer $WS_2$. Versatile light emission control along all device orientations is achieved for different emitters and excitation wavelengths, benefiting from the facile size control and isotropic shape of SiNSs. Simultaneous modulation of excitation and emission via a single SiNS at visible wavelengths significantly improves the efficiency and directivity of TMD exciton emission and leads to the potential of multifunctional integrated photonics. Overall, our work opens promising opportunities for nanophotonics and polaritonic systems, enabling efficient manipulation, enhancement and reconfigurability of light-matter interactions.

Recent advances in two-dimensional (2D) semiconductors[1-4], quantum dots[5] and color centers[6] have showcased several opportunities for next-generation integrated photonic devices, such as nanoscale light sources. Effective control over the emission properties is of great importance for these novel emitters in order to realize optimized functionalities. For instance, the low quantum efficiency and poor emission directivity of monolayer transition-metal dichalcogenides (TMDs) have limited their practical applications in integrated photonics[7] and flexible optoelectronic systems[8] due to unwanted signal degradation. Tailoring the incident and radiation fields is thus particularly meaningful. However, challenges exist in both efficiency and complexity. When optical components are coupled at the nanoscale, not only material losses cause decreased efficiency, but also very fine alignment between elements is required. Therefore, a single subwavelength modulator working at visible wavelengths is highly desired for achieving higher coupling efficiency and miniaturized device size.

The last decades have witnessed the rapid development of optical nanoantennas as a promising solution to manipulating optical fields at the nanoscale[9-12], leading to drastically enhanced light emission[10,13,14], photodetection[15] and optical sensing[16]. Both plasmonic resonances based on noble metals[17-20] and Mie resonances from high-index dielectric materials[21-29] have been explored to facilitate strong light-matter interactions in the near field. Uniquely, dielectric nanoantennas allow simultaneous excitation of magnetic and electric resonances[28,30], whereas generating magnetic responses at optical frequencies may be challenging in plasmonic structures[29,31-33]. Dielectric nanoantennas can therefore implement interesting optical field manipulations within extremely simple geometries[34-36], e.g., a single nanosphere, as compared to complex shapes and arrays typically required in plasmonic designs[37]. In addition, metals fundamentally suffer higher material losses than dielectrics, especially in the visible region[38], which may further decrease the already low efficiency of nanoemitters. Consequently, the use of dielectric resonators appears to be an ideal solution to develop efficient subwavelength emission modulators.

The mutual interference of size-dependent magnetic and electric modes in dielectric nanoantennas can be used to efficiently modulate the far-field radiation pattern[28,39]. For example, directional scattering of plane waves has been readily achieved at microwave[40], THz[41], and optical frequencies[42-44], combining electric and magnetic resonances at the Kerker conditions[23,45,46]. Since coupling with nanoantennas can significantly modify the emission properties of emitters via the Purcell effect and sophisticated multipolar interference, positioning suitably designed dielectric resonators close to an emitter is a potential route to achieve highly directional emission[43,47-49]. Moreover, as demonstrated in the plasmonic regime[37] and theoretically proposed for Mie antennas[47], a combination of directivity modulation of both excitation and emission can provide more degrees of freedom to control the overall emission process. Cihan *et al.* have demonstrated directivity modulation of monolayer $MoS_2$ emission with silicon (Si) nanowires[50]. But the challenge on the device size and the inconsistency in modulation depth along different device orientations[65, 66] (i.e., the radial and axial orientations of nanowires) remain, limiting the future device integration. From this perspective, dielectric nanospheres represent a compelling platform: on the one hand, thanks to their subwavelength nature in all dimensions, they can strongly enhance the emission of coupled emitters along all sample orientations; on the other hand, they also provide smaller footprints as well as effective trapping of the excitation light, which both exerts the advantage of compactness and remedies the low efficiency. Finally, given their mature industrial base, specifically Si-based nanoantennas[21] can ensure better compatibility with existing CMOS and emerging integrated photonic platforms[51].

In the literature, the controllable emission from a dipole coupled with a dielectric sphere has been demonstrated at microwaves[52]. However, it is still challenging to practically realize such a subwavelength platform in the visible region using silicon nanospheres. To this extent, a universal analytical model for three-dimensional (3D) systems[48,50] can provide physical insights into how the dipole-excited Mie resonances interfere and describe the evolution of the radiation patterns under various conditions. Sphere size control and good sphericity are also important, but can hardly be realized[42,53], especially when lower-order modes (smaller sphere size) are required at visible wavelengths[22,24,28].

In this work, we resolve these challenges by proposing a highly miniaturized emission control platform based on single subwavelength nanospheres. A rigorous multipolar model is first derived based on Mie theory, in 3D, to describe the far-field radiation pattern of the nanosphere-modulated dipole emission. Based on reciprocity theorem, this model instructs the possibility and optimal conditions of directivity modulation of both excitation and emission processes via single nanospheres (Fig. 1a, b). Then, we experimentally demonstrate versatile directivity modulation of 532 nm laser excitation and 635 nm exciton emission from a monolayer $WS_2$ with controllable forward-to-backward intensity (F/B) ratios in 125 samples via spherical dielectric nanoresonators, showing statistical agreement with our theoretical model and numerical simulations. The employed nanoantennas are single hydrogenated amorphous Si nanospheres (a-SiNS:Hs) with facile size control (200 to 500 nm)[54] and excellent sphericity (see Fig. S1). The a-SiNS:Hs support low-loss multipolar resonances down to 450 nm wavelength, spanning the whole visible range[55]. Unless otherwise noted, we will refer a-SiNS:Hs as SiNSs for the ease of reading. Based on this platform we tailor with large flexibility the emission properties. Various emitters and excitation wavelengths are tested to demonstrate highly directional forward emission with maximized forward excitation efficiency, matching well with the predicted performance. Under optimized directional excitation, highly directional emission with a total enhancement up to 5 folds is observed, which significantly enhances the efficiency and directivity of the emission by 2D TMDs or other nanoemitters. Integration of TMDs to resonant nanostructures has the advantage of controllable and accurate assembly and low-cost fabrication[1,2,50]. Our results manifest the efficient and versatile modulation of exciton emission at visible wavelengths via a single subwavelength nanosphere and thus promote the device miniaturization in all dimensions, opening promising opportunities for nanophotonics[2], valleytronics[56] and polaritonic systems[57].

**Analytical theory and numerical simulations**

While the optical properties of nanoantennas can be evaluated numerically, it is of both theoretical and practical interest to perform an analytical study to gain physical insights into the role of each resonance in the total directivity modulation. For the sake of convenience, we omit the presence of a substrate in this analysis and consider a SiNS with diameter $2a$ coupled with a tangential electric dipole emitter positioned at the distance $d = 1$ nm. The localized dipole models the exciton emission from the monolayer TMD. Because of the Mie resonances supported by the SiNS, this configuration is able to redistribute both excitation power on the dipole[47] and the emitted power from the dipole[28,48,49] for different conditions, resulting in tunable directional excitation and emission. The directivity modulations of excitation and emission processes are reciprocal of each other. (See Fig. 1a, b and Supporting Information (SI) Section IV)

The scattering problem of a sphere excited by a coupled electric dipole has been analyzed in the literature but mainly in the near-field region to study the modification of the emitter's decay rates[58,59]. In order to quantify the directivity, we revisit this problem in the far-field region. Following a similar procedure as in standard Lorenz-Mie theory, we expand the incident field, i.e., the exciton emission, and the scattered field, into spherical vector harmonics[60,61]. Compared with a recent work on nanowires[50], the three-dimensional (3D) nature of the present system introduces significant complexity in the algebra. For a tangential electric dipole located at $(a + d, 0, 0)$ in spherical coordinates, the entire field can be derived from Debye potentials $u$ and $v$, which are connected to the auxiliary vertical electric and magnetic dipole moments, respectively. The final expression of the potentials reads:

$$u = \frac{i}{r_s \cdot r} \sin\theta \cos\varphi \sum_{n=1}^{\infty} \frac{(2n+1)}{n(n+1)} [a_n \zeta'_n(k_0 r_s) - \varphi'_n(k_0 r_s)] \zeta_n(k_0 r) P'_n(\cos\theta) \qquad (1)$$

and

$$v = -\frac{1}{r_s \cdot r} \sin\theta \sin\varphi \sum_{n=1}^{\infty} \frac{(2n+1)}{n(n+1)} [b_n \zeta_n(k_0 r_s) - \varphi_n(k_0 r_s)] \zeta_n(k_0 r) P'_n(\cos\theta) \qquad (2)$$

for the region $r > r_s = a + d$. Here, $a_n$ and $b_n$ are the Mie coefficients, known as functions of the size parameter $k \cdot a$; $k_0$ and $k$ are the wave vectors in free space and in SiNS, respectively; $\varphi_n$ and $\zeta_n$ are the Riccati-Bessel functions related to the spherical Bessel function and spherical Hankel function of the first kind, respectively; $P_n$ is the Legendre polynomials and the prime denotes a derivative. Details of the derivation can be found in SI Section III. The dependence of the Mie coefficients on $k \cdot a$ suggests that the size parameter plays an important role in the directivity modulation performance, showing the potential for versatile designs in both excitation and emission wavelengths via SiNS sizes.

Utilizing the developed model, we first examine the radiation patterns for different parameters by evaluating the outgoing Poynting vector in the far-field region. The dielectric function of Si is confirmed by fitting the scattering spectra as discussed in SI Section I. In Fig. 1a and b, two examples are presented for SiNSs of different radii and at different emission wavelengths. A highly forward directed modulation can be found at 532 nm wavelength with a 390 nm SiNS (Fig. 1a), while more backward components can also be achieved at 635 nm wavelength with a 250 nm SiNS (Fig. 1b). Importantly, reciprocity (SI Section IV) ensures that the near field intensity at the position of the emitter excited by a 532 nm plane wave will be similarly modulated by a 390 nm SiNS, with the same F/B ratio. 532 and 635 nm wavelengths are chosen for illustration according to the excitation and emission studied experimentally in the next section.

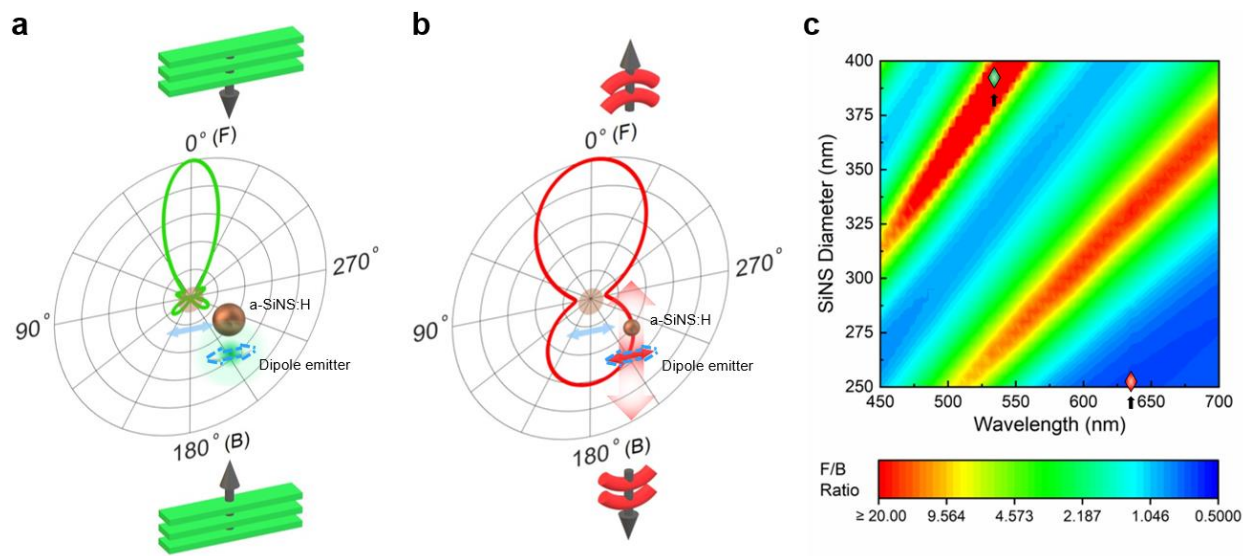

**Figure 1. Directivity control on both excitation and emission processes of a dipole via single SiNS resonators. a, b.** Analytical calculation of the radiation patterns of **a** a 532 nm (wavelength) dipole coupled to a 390 nm (diameter) SiNS and **b** a 635 nm (wavelength) dipole coupled to a 250 nm (diameter) SiNS. 0 degrees is defined as forward (F), while 180 degrees is defined as backward (B). Considering the reciprocity theorem (See Fig. S5), the extracted forward-to-backward (F/B) ratio of the emission process (schematic in **b**) can also be used to describe the forward and backward excitation of the near field at the position of the dipole (schematic in **a**). **c.** Numerical simulation of the F/B ratio mapping as a function of wavelength and SiNS diameter. A glass substrate is considered as compared to the free-standing SiNS in the analytical model. The corresponding cases in **a** and **b** are denoted by the green and red rhombi, respectively. Influence of the glass substrate is presented in Fig. S7.

To account for the substrate effect and make the model better suited to fit the experimental demonstration, we also conduct full-wave simulations with the presence of a semi-infinite substrate of glass. Fig. 1c shows that the ratio of incident/emitted power in the forward direction to the one in the backward direction (F/B ratio), as a function of wavelength and the size of SiNS. The mapping by the analytical approach shows basically the same tendency in Fig. S6. Since the emission property is determined by both dipole excitation and its decay channels, the modulation is two-fold: on the one hand, the nanoemitter needs to be coupled with an SiNS of optimal size to realize the most efficient excitation at a given wavelength and from a certain direction; on the other hand, the SiNS should be tailored to also enhance radiation in the preferred direction at the emission wavelength. Here, figures 1c and S6 can serve as a graphical guide to determine the dimension of SiNS, excitation wavelength, and incident direction for implementing the desired emission properties.

**Controllable directivity modulation of excitation and emission of WS₂ excitons**

To experimentally demonstrate controllable F/B ratios, a CVD-grown monolayer WS₂ flake is selected as the emitter (see SI Section II), and SiNSs are drop casted on the top, as shown in Fig. 2a. We modify our microscope system with laser excitation from both top and bottom (Fig. 2b) to support forward and backward modulated excitation/emission when the sample faces either up or down. In order to study the directivity modulation of the excitation and emission individually, we have to separate these two processes efficiently:

$$I_{F(B)}^{Ex(Em)} = f_{F(B)}(I_0^{Ex(Em)}, P_0) \cong \alpha_{F(B)}^{Ex(Em)} \cdot I_0^{Ex(Em)}, \qquad (3)$$

Where $P_0$ is the incident power density, $I_0^{Ex(Em)}$ is the excitation(Ex)/emission(Em) light intensity without modulation, $I_{F(B)}^{Ex(Em)}$ is the forward (backward) modulated intensity via SiNS. A constant $\alpha_{F(B)}^{Ex(Em)}$ can be used to represent the modulation function $f_{F(B)}$ in the low power (linear) regime ($P_0 \leq 70$ µWµm$^{-2}$). Experimental proofs and detailed discussions can be found in SI Section VII.

The comparison of the forward and backward modulated emission is carried out when the modulation of excitation is along a fixed direction, and vice versa for studying the directivity of excitation. For example, as shown in Fig. 2b, with our sample facing down we collect the forward modulated emission signals from the bottom and determine the F/B ratio for excitation by comparing the collected signal intensities under forward and backward excitation conditions. More details on the measurements can be found in SI Section VIII.

Finally, the results from 125 samples are summarized in Fig. 2c and d, showing a good agreement with our theoretical predictions statistically. They reveal that controllable directivity modulation can be simultaneously achieved in both excitation and emission. The combined effect can thus be designed to tailor the overall exciton emission properties, given that the size of SiNSs can be well controlled across several hundreds of nanometers. It is worth noting that the directivity can become much larger if the part of monolayer WS₂ not covered by the SiNS is etched[45] or a smaller N.A. is used for signal collection. That is why we see an obvious difference in F/B ratios between experiments and simulations when it is >1. As for the relatively similar F/B ratios when it is <1 and other discussions, please see the detailed explanations in SI Section VIII. Anyhow, this work mainly focuses on controllable directivity, and the current results in Fig. 2c,d clearly demonstrate the phenomenon. From this perspective, we extract the system-dependent relation (Eq. S12) between simulated and measured F/B values (see *y*-axes in Fig. 2c, d) and apply it to predict the performance in more versatile emission designs in the following.

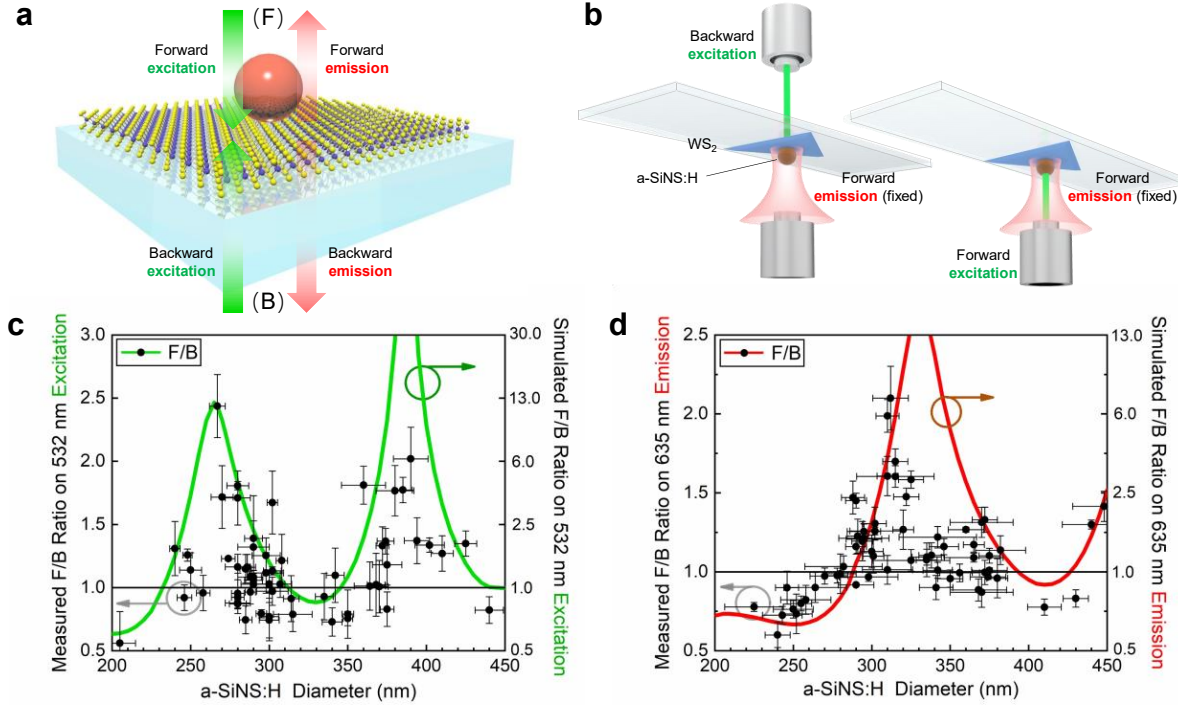

**Figure 2. Schematic diagram and experimental demonstration of controllable directivity modulations of excitation and emission separately. a.** Schematic of the monolayer WS$_2$ emitter modulated by a single SiNS on it. F: forward. B: backward. **b.** Sketch of the experimental setups measuring the forward modulated WS$_2$ emission (sample facing down) under forward excitation (Right) and backward excitation (Left), respectively. Backward modulated emission can be measured similarly with sample facing up. The blue triangles refer to CVD-grown monolayer WS$_2$ flakes. **c, d.** Measured and simulated F/B ratios as a function of SiNS diameter on both **c** 532 nm excitation and **d** 635 nm WS$_2$ emission. The ranges of *y*-axes are selected for better display. See the whole curves for simulated F/B ratios in Fig. S10.

The F/B ratio can be up to 2.5 and down to 0.5 owing to the mutual interference of different resonances supported by SiNS. The mode contributions at different wavelengths can be extracted from our modified Mie theory for dipole excitation, yielding the scattering efficiency

$$Q_{dipole} = \frac{3}{4k_0^2 r_s^2} \sum_{n=1}^{\infty} (2n+1)\left[|a_n|^2 |\zeta'_n(k_0 r_s)|^2 + |b_n|^2 |\zeta_n(k_0 r_s)|^2\right]. \quad (4)$$

Compared to the standard model for plane-wave excitation, $a_n$ and $b_n$ here are modulated by the spherical harmonics $\zeta_n$ and their derivatives, respectively. Two examples are presented in Fig. 3a, c. For fixed wavelength, the multipolar superposition at different SiNS sizes is also drawn in the shaded background of Fig. S10 for both 532 nm excitation and 635 nm emission. Figure 3b illustrates the forward-enhanced 532 nm excitation enabled by a 390 nm SiNS, as analytically studied in Fig. 1a. As another typical scenario of interest highlighted in Fig. 1b, backward-enhanced 635 nm emission enabled by a 250 nm SiNS is shown in Fig. 3d. All these results are normalized based on the isolated monolayer $WS_2$ emission without modulation (black curves in Fig. 3b, d). For a given SiNS modulator, we can intuitively understand the directivity modulation as the difference between the solid (forward) and dashed (backward) colored curves. However, in Fig. 3b and d, each emission spectrum represents the total modulation of the emission property, instead of the separate modulation of excitation and emission discussed so far. For instance, the forward enhanced 532 nm excitation via a 250 nm SiNS (Fig. 3d) makes the total emission always larger than the pure $WS_2$ emission no matter whether it emits in the forward or backward direction. Similarly, we can find that the green curves are both larger than the black curve in Fig. 3b, due to the backward enhanced modulation of emission. The schematics of the two-step modulation according to Fig. 3b, d can be found in Fig. S12.

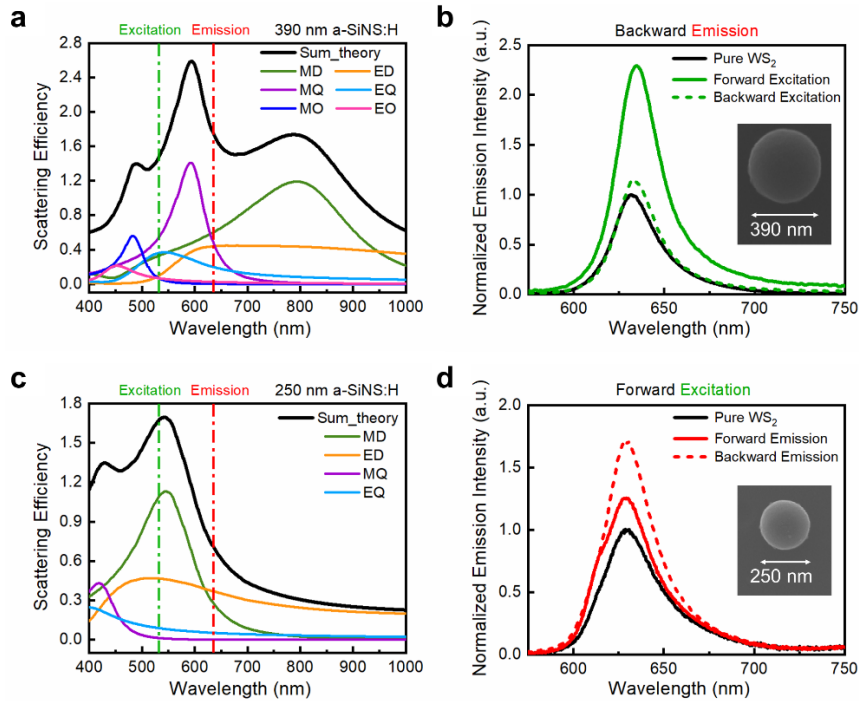

**Figure 3. Interference between multipoles and simultaneous modulation of excitation and emission for total light emission control. a.** Scattering efficiency and its multipolar contributions of a 390 nm SiNS based on the Mie theory modified for dipole excitation. The green and red vertical dashed lines denote the excitation and emission wavelengths, respectively. **b.** The backward emission from monolayer $WS_2$ under forward and backward excitation condition, modulated by a 390 nm SiNS. **c.** The same as **a**, but with a 250

nm SiNS. **d.** The forward-excited emission from monolayer WS$_2$ collected in the forward and backward directions, modulated by a 250 nm SiNS. All the curves in **b** and **d** are normalized to pure WS$_2$ emission without any modulation. MD: magnetic dipole, ED: electric dipole, MQ: magnetic quadrupole, EQ: electric quadrupole, MO: magnetic octupole, EO: electric octupole.

In order to study the directivity modulation exclusively on the excitation (emission) process, the modulation of emission (excitation) must be fixed along a specific direction, either forward or backward. This provides us two choices in experiments, and we can choose the fixed direction that gives us a higher signal-to-noise ratio in the measurements (e.g., stronger peaks in green and red in Fig. 3b, d). The mechanism behind such a choice can be attributed to the different magnitude and phase of the multipoles in a SiNS when excited by a 532 nm plane wave or a 635 nm dipole emission. The field intensities at arbitrary positions modulated by these multipolar resonances can be easily extracted from our multipolar derivations and straightforwardly displayed in phasor diagrams, for instance as shown in Fig. 4. Then, thanks to the universality of our model and the control on the SiNS size, the phasor diagrams can provide us the optimal parameters for emission control at any excitation wavelength.

**Versatile emission control of emitters**

To demonstrate our ability of versatile emission control, we start from the design of highly directional forward emission under maximum forward excitation efficiency by choosing suitable SiNSs for different emitters (monolayer WS$_2$ or MoS$_2$) and excitation wavelengths (446 or 532 nm). Figure 4a, b and d, e present the basic logic of the design for a minimum backward component in both emission and excitation intensity distributions. Here, the maximum forward excitation is simply chosen for easier experimental demonstrations, while backward excitation can also be designed straightforwardly. Although we derive the complex values for each phasor vector based on a common plane-wave-sphere model (Fig. 4a, d) and a modified dipole-sphere model (Fig. 4b, e) respectively, the reciprocity theorem still works for every decomposed variable (See SI Section XI).

Thanks to the multipolar resonances, the incident field (*INC*) can be almost cancelled out along the backward direction, as shown in the rational designs of 446 nm excited WS$_2$ via a 320 nm SiNS (Fig. 4a) and 532 nm excited MoS$_2$ via a 385 nm SiNS (Fig. 4d). Meanwhile, the forward direction shows significant enhancement in total as illustrated in Fig. S13a, c. Based on the same SiNS, good forward directivity is also achieved for emission as shown in Fig. 4b, e and S13b, d. Comparing Fig. 4a(d) and b(e), we notice that fewer higher-order modes are needed for emission phasor diagrams but still with enough design accuracies. This is because of the longer wavelengths of emission than that of excitation. For a certain SiNS size, fewer resonances can be effectively triggered by photons with smaller energies. As another proof, changes in the radiation patterns can hardly be observed when additional higher-order modes are included in analysis, as shown in Fig. S4, in agreement with our analysis.

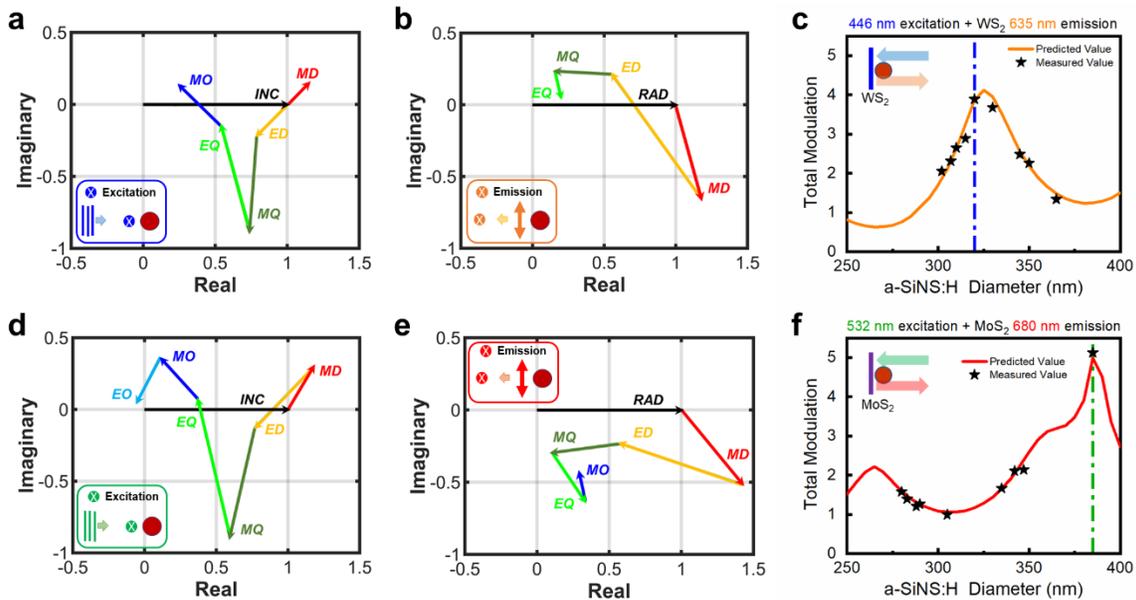

**Figure 4. Achieving targeted emission properties for different emitters and excitation wavelengths via SiNS sizes. a, b.** Phasor diagrams showing the backward modulation of both **a** 446 nm excitation and **b** 635 nm WS$_2$ emission via a 320 nm SiNS. The cross points in the insets denote the positions where the phasor is extracted. The incident field (INC), radiation field (RAD), resonant MD, ED, MQ, EQ, MO, and EO are labelled for each vector. All of them are normalized based on INC/RAD. The minimum backward components in **a, b** lead to the design of highly directional forward emission under maximum forward excitation efficiency, as highlighted by the blue dashed line in **c**. **c.** Predicted and measured total modulation (combination of directivity modulations of excitation and emission) against SiNS sizes. **d, e.** The same design as **a, b** for 532 nm excitation and 680 nm MoS$_2$ emission via a 385 nm SiNS, and **f.** their corresponding prediction accuracy. Typical emission spectra of the measured data points in **c, f** can be found in Fig. S14.

Through the same mechanism, we can predict the overall modulation under any condition (i.e., different emitters, excitations, directions, and SiNS sizes). Figure 4c and f show the corresponding prediction for forward excitation plus forward emission. Total modulation up to 5-fold and down to 1-fold are found, and good agreement with our theoretical model is observed, showing a solid performance of our SiNS-based platform. A video (SI Section XIII) is also provided to show the emission control performance in real time.

Similarly, maximized highly directional backward emission or totally suppressed emission with low excitation efficiency can be accurately designed. Interestingly, the same SiNS-emitter hybrid can perform different functions by tuning the excitation wavelengths under rational designs, as shown in Fig. S15. This provides the two degrees of freedoms of SiNS size and excitation wavelength to fit the desired emitter performance, leading to the potential of ultra-compact and multiplexed integrated photonics.

**Conclusions**

In summary, we have experimentally demonstrated effective control of exciton emission via a low-loss subwavelength SiNS at visible wavelengths through controllable directivity modulation of both excitation and emission processes. The isotropic spherical shape allows consistent directivity modulation along all sample orientations. Based on a modified Mie theory for dipole excitation, the control over F/B ratio is attributed to the superposition of multipolar resonances supported by single SiNSs. Phasor diagrams are extracted from the analytical model and provide great insights into the size and wavelength dependent multipolar contributions. Measurements performed on 125 SiNS resonators convincingly suggest a good performance of our highly miniaturized platform and highlight the two-step directivity modulation of incident and radiation fields. Finally, thanks to the universality of our model and the facile SiNS size control, we achieved versatile emission property designs on various emitters. Two degrees of freedoms of SiNS size and excitation wavelength in the design provide us opportunities for multi-functional nanophotonics. Moreover, with the rigorous consideration of the emitter at various position/orientation in derivations[62], our multipolar theory also allows for the analysis of the dipole oriented normal to the sphere's surface, e.g., spin-forbidden dark exciton emission in TMDs[63].

Our work boosts the development of ultra-compact and multiplexed integrated photonic devices at visible wavelengths by a silicon-based subwavelength nanoantenna. As the diffraction limit in our measurements significantly degrades the detected signals in directivity, we would expect enhanced performance when our platform is incorporated in silicon nanophotonics, e.g., nanowaveguides[51]. On-demand modulator assembly for the wanted functions might also be realized by a size-selective optical printing[64] based on our SiNSs. The proposed antenna-emitter hybrid may give insights into the use of high-index dielectric nanoparticles as functional components in photonics circuits.

**Methods**

Methods are available at Supplementary Information.

**Acknowledgements**

The authors would like to thank S. Lepeshov for his help on full-field simulations, and D. Kim for his help on emission spectra measurements. J.F., K.Y. and Y.Z. acknowledge the financial support of the National Aeronautics and Space Administration Early Career Faculty Award (80NSSC17K0520), the National Science Foundation (NSF-CBET-1704634 and NSFCMMI-1761743), and the National Institute of General Medical Sciences of the National Institutes of Health (DP2GM128446). M.W. acknowledges the financial support of University Graduate Continuing Fellowship of the University of Texas at Austin. M.W., A.K. and A.A. acknowledge the financial support of the Air Force Office of Scientific Research, the Department of Defense, the Simons Foundation and the National Science Foundation. T.J. and B.A.K. acknowledge the financial support of the Robert A. Welch Foundation (F-1464) and the National Science Foundation through the Center for Dynamics and Control of Materials (CDCM) Materials Research Science and Engineering Center (MRSEC) (DMR-1720595). Partial support for J.C. was provided by the Department of Energy, Basic Energy Science program via grant DE-SC0019398 and X.L. gratefully acknowledge the Welch foundation via grant F-1662. T.Z., E.K. and M.T. acknowledge the financial support of the Air Force Office of Scientific Research (AFOSR) through grant No. FA9550-18-1-0072.

**Author contributions**

Y.Z., A.A., M.W., J.F. and K.Y. conceived and coordinated the project. J.F. and M.W. designed the experiments. J.F., M.W., J.C., and X.L. carried out the experiments. K.Y. conducted the theoretical derivations. K.Y., J.F. and A.K. conducted full-field simulations. T.Z., E.K. and M.T. prepared the TMDs. T.J. and B.A.K. synthesized the a-SiNS:H. J.F., M.W. and K.Y. wrote the manuscript. All the authors discussed the results and commented on the manuscript.

**Competing interests**

# Supplementary Information:

# Directivity modulation of exciton emission using single dielectric nanospheres

Jie Fang[1,*], Mingsong Wang[1,2,*✉], Kan Yao[1,*], Tianyi Zhang[3], Alex Krasnok[2], Taizhi Jiang[4], Junho Choi[5], Ethan Kahn[3], Brian A. Korgel[4], Mauricio Terrones[3], Xiaoqin Li[5], Andrea Alù[2✉], and Yuebing Zheng[1✉]

[1]Walker Department of Mechanical Engineering and Texas Materials Institute, The University of Texas at Austin, Austin, TX 78712, USA

[2]Photonics Initiative, Advanced Science Research Center, City University of New York, New York, NY 10031, USA

[3]Department of Materials Science and Engineering, Department of Physics, Department of Chemistry and Center for 2-Dimensional and Layered Materials, The Pennsylvania State University, University Park, PA 16802, USA

[4]McKetta Department of Chemical Engineering, The University of Texas at Austin, Austin, TX 78712, USA.

[5]Department of Physics, The University of Texas at Austin, Austin, TX 78712, USA

*These authors contributed equally: Jie Fang, Mingsong Wang, Kan Yao

✉Email: mwang2@gc.cuny.edu, aalu@gc.cuny.edu, zheng@austin.utexas.edu

**I. Synthesis and characterization of a-SiNS:Hs**

**II. Preparation and characterization of monolayer WS$_2$**

**III. Derivation of the analytical multipolar model**

**IV. Reciprocity theorem and simulations**

**V. Mapping of F/B ratio by analytical model**

**VI. Influence of the glass substrate on radiation patterns**

**VII. Low power approximation to effectively separate the modulation of excitation and emission**

**VIII. Measurements and data analysis for F/B ratios**

**IX. Converting the simulated F/B ratios into measured (predicted) values**

**X. Total modulation of both excitation and emission**

**XI. Reciprocity theorem in phasor diagrams**

**XII. Supplementary phasor diagrams**

**XIII. Emission control on different emitters and excitation wavelengths**

**XIV. Ultra-compact and multiplexed integrated photonics**

## I. Synthesis and characterization of a-SiNS:Hs

*Synthesis:* A 10 mL titanium batch reactor (High-Pressure Equipment Company (HiP Co.) is used for the synthesis. First, 21 µL trisilane ($Si_3H_8$, 100%, Voltaix) and n-hexane (anhydrous, 95%, Sigma-Aldrich) are loaded in the reactor in a nitrogen-filled glovebox. The amount of n-hexane loaded in the reactor is associated with the reaction pressure inside the reactor during the heating process. The hydrogen concentration in a-SiNS:Hs is determined by different reaction temperatures[1]. For example, a-SiNS:H with a hydrogen concentration of 40% is synthesized at a temperature of 380 °C and a pressure of 34.5 MPa (5000 psi). After adding the reagents, the reactor is sealed by using a wrench inside the glove box. Then a vice is used to tightly seal the reactor after removing it from the glove box. The reactor is heated to the target temperature in a heating block for 10 min to allow the complete decomposition of trisilane. After the reaction, an ice bath is used to cool the reactor to room temperature. Colloidal a-SiNS:Hs are then extracted from the opened reactor. The a-SiNS:Hs are washed by chloroform (99.9%, Sigma-Aldrich) using a centrifuge (at 8000 rpm for 5 min).

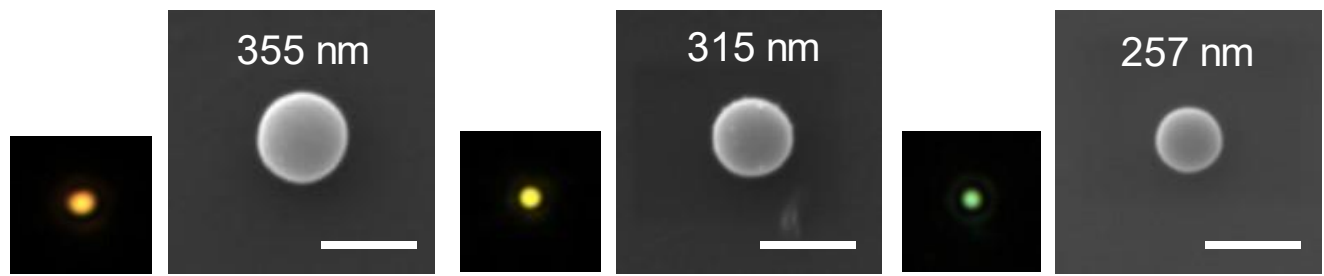

**Figure S1. Scattering and SEM images of a-SiNS:Hs** showing flexible size control, good sphericity, and tunable resonances at visible wavelengths. Scale bar: 350 nm.

*Characterization:* The size of a-SiNS:H is measured by SEM as shown in Fig. S1. The reaction temperature and pressure together determine the range of sizes for synthesized a-SiNS:H. The a-SiNS:H shows flexible size control, good sphericity, and tunable resonances at visible wavelengths (See their dark-field scattering images in Fig. S1).

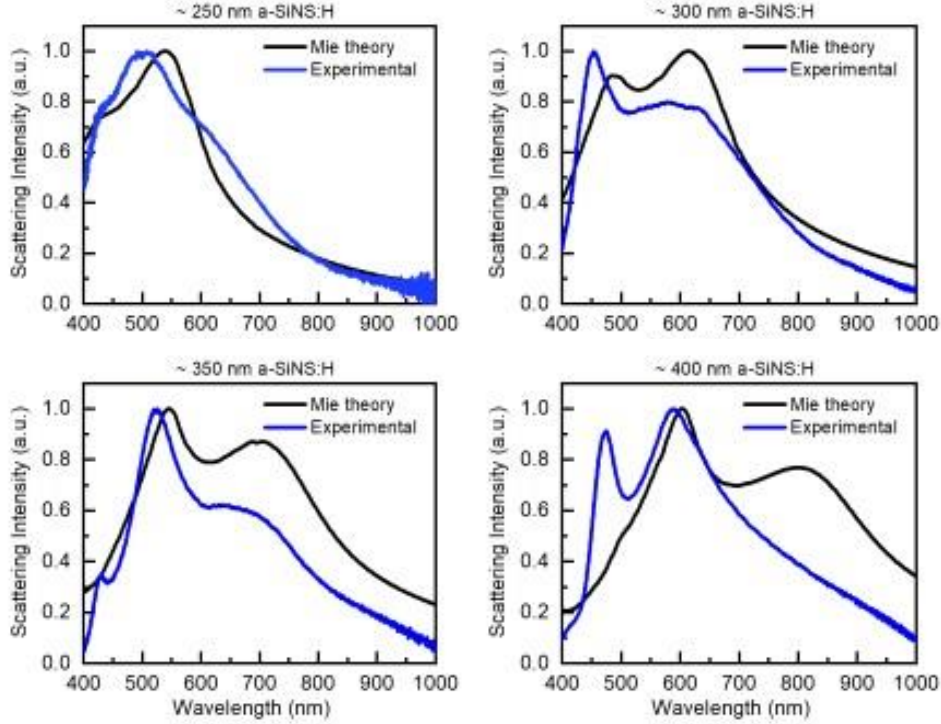

**Figure S2. Fitting the dielectric function of a-SiNS:Hs.** Examples of the well-fitted scattering spectra based on the same dielectric function for around 250 nm (left-top), 300 nm (right-top), 350 nm (left-bottom), and 400 nm (right-bottom) a-SiNS:Hs.

As discussed in our previous work[2], the voids in a-SiNS:H brought by hydrogen atoms cause a distortion of Si-Si bond, which leads to different bonding angles, smaller Si-Si distances and thus a larger bandgap, compared to that for crystalline Si. It should be noted that, for the ease of experimental demonstration and versatile emission control based on our platform, the same hydrogen concentration is chosen for all the a-SiNS:Hs of different sizes in this work. Their transmission spectra in Ref.[2] show the lossless nature down to 450 nm in wavelength.

To model the permittivity of the a-SiNS:H in use, we apply the Maxwell-Garnett mixing formula

$$\varepsilon_{a-SiNS:H} = \varepsilon_{Si} \frac{(\varepsilon_{voids}+2\varepsilon_{Si})+2f(\varepsilon_{voids}-\varepsilon_{Si})}{(\varepsilon_{voids}+2\varepsilon_{Si})-f(\varepsilon_{voids}-\varepsilon_{Si})}, \quad (S1)$$

where $f$ is the volume fraction of hydrogenated voids and $\varepsilon_{voids}$ denotes the permittivity of voids, which is assumed to be 1. We also slightly blue-shift the permittivity after such an ideal mixture (no distortion considered)[3] to model the hydrogenation-induced bandgap renormalization. By using $f = 0.75$ and a blue-shift of 150 nm, we obtain the best matching of the calculated scattering peaks (by Mie theory) with experimental data of all the a-SiNS:Hs in use. Four examples are shown in Fig. S2.

## II. Preparation and characterization of monolayer WS₂

*Synthesis:* Monolayer WS$_2$ is synthesized on SiO$_2$/Si by an atmospheric-pressure chemical vapor deposition (APCVD) method[4]. First, powders of WO$_3$ (5 mg) and NaBr (0.5 mg) are mixed and placed on a piece of SiO$_2$/Si wafer in an alumina boat. Then another piece of SiO$_2$/Si (serving as the growth substrate) is placed on the top of the alumina boat facing down and loaded inside a quartz tube. Sulfur powder (400 mg) is loaded in another alumina boat on the upstream. The furnace is ramped up to 825 °C and held for 15 min during synthesis, and sulfur powder is evaporated at 250 °C separately using a heating belt. 100 sccm of argon is used as the carrier gas.

*Transfer:* As-synthesized monolayer WS$_2$ is transferred onto glass coverslips (0.17 mm thick) via a cellulose acetate (CA)-based wet transfer method that we have described previously[5]. The use of coverslips instead of microscope slides enables us to face the sample either up or down, but never beyond the working distances of the objectives.

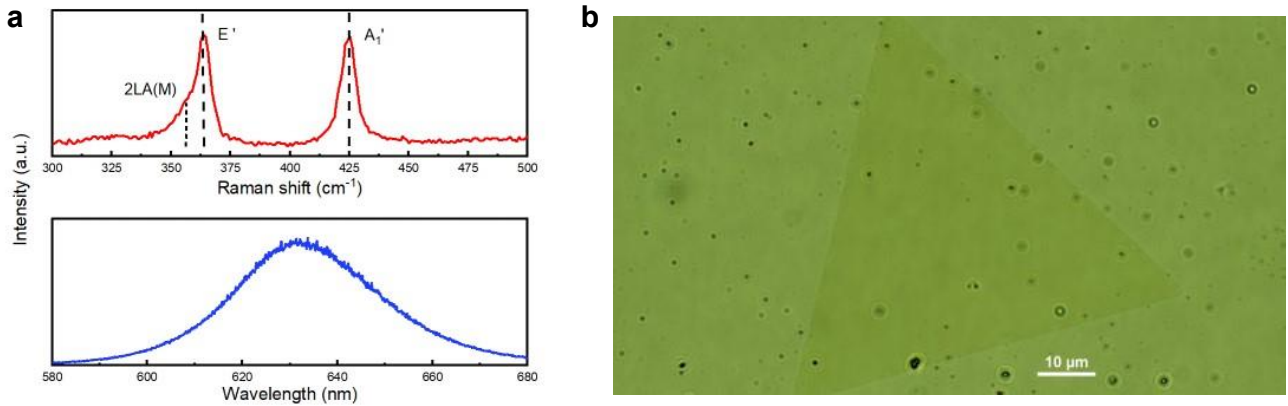

**Figure S3. Characterization of monolayer 1H-WS$_2$. a** Raman (top) and photoluminescence (bottom) spectra for monolayer WS$_2$ without a-SiNS:Hs. **b** Optical image showing a-SiNS:Hs of different sizes (modulation functions) drop casted on a monolayer WS$_2$ flake.

*Characterization:* The monolayer nature of the CVD-grown WS$_2$ flake is confirmed by the strong photoluminescence peak at ~635 nm (1.95 eV)[6] (Fig. S3a, bottom panel). We also examine the degree of crystallinity of the monolayer WS$_2$ by measuring the Raman scattering spectrum with a 488 nm excitation laser (Fig. S3a, top panel). The Raman spectrum is dominated by three peaks at 355, 363, and 425 cm$^{-1}$, which correspond to the second-order longitudinal acoustic 2LA(M) mode, the first-order out-of-plane E' mode, and the first-order in-plane A$_1$' mode of monolayer WS$_2$, respectively[7,8]. Through the optical contrast, we can clearly see a representative WS$_2$ flake on the glass coverslip in Fig. S3b, with a-SiNS:Hs of different sizes (modulation functions) drop casted on it.

## III. Derivation of the analytical multipolar model

This section presents the derivation of the analytical model for the emitter-sphere hybrid. Similar configurations have been explored previously in the context of near-field behaviors, such as radio-waves propagating along the surface of the earth[9] and the modification of decay rates of molecules near a particle[10-13]. Here, we revisit this problem with a different focus on how the dipole-excited resonances of the a-SiNS:H will affect the angular distribution of the emission in the far-field. We consider a sphere of radius *a* located at the origin of the spherical coordinate system ($r$, $\theta$, $\varphi$), and the light source, a point electric dipole oriented along the *x*-axis, is positioned at ($a + d$, 0, 0). In other words, the dipole is at a distance *d* above the north pole of the sphere. A convenient way to solve scattering problems involving a sphere is to decompose the total fields into two components, which, respectively, have the electric and magnetic fields transverse to the radial direction, known as the Derby potentials *u* and *v*. These auxiliary potentials can be expressed by infinite sums of spherical wave functions as:

$$u = \frac{i}{r_s \cdot r} \sin\theta \cos\varphi \sum_{n=1}^{\infty} \frac{(2n+1)}{n(n+1)} [a_n \zeta_n'(k_0 r_s) - \varphi_n'(k_0 r_s)] \zeta_n(k_0 r) P_n'(\cos\theta) \quad \text{(S2)}$$

and

$$v = -\frac{1}{r_s \cdot r} \sin\theta \sin\varphi \sum_{n=1}^{\infty} \frac{(2n+1)}{n(n+1)} [b_n \zeta_n(k_0 r_s) - \varphi_n(k_0 r_s)] \zeta_n(k_0 r) P_n'(\cos\theta) \quad \text{(S3)}$$

for the region $r > r_s = a + d$, with $a_n$ and $b_n$ the Mie coefficients given by

$$a_n = \frac{k \varphi_n'(k_0 a) \varphi_n(ka) - k_0 \varphi_n(k_0 a) \varphi_n'(ka)}{k \zeta_n'(k_0 a) \varphi_n(ka) - k_0 \zeta_n(k_0 a) \varphi_n'(ka)} \quad \text{(S4)}$$

and

$$b_n = \frac{k \varphi_n(k_0 a) \varphi_n'(ka) - k_0 \varphi_n'(k_0 a) \varphi_n(ka)}{k \zeta_n(k_0 a) \varphi_n'(ka) - k_0 \zeta_n'(k_0 a) \varphi_n(ka)}. \quad \text{(S5)}$$

Here, $k_0$ and $k$ are the wave vectors in free space and in the sphere, respectively; $\varphi_n$ and $\zeta_n$ are the Riccati-Bessel functions related to the spherical Bessel function and spherical Hankel function of the first kind, respectively; $P_n$ is the Legendre polynomials and the prime denotes a derivative. Note that *u* and *v* are both composed of two parts as seen in the square brackets. The terms containing a Mie coefficient represent the scattered fields by the sphere in response to the incident fields from the dipole, and the other ones without a coefficient are the outgoing waves directly from the dipole emission. Because the radiation pattern is given by the outgoing power at a constant radius in the far-field, we only need to know the electric and magnetic fields in the polar and azimuthal directions to compute the radial component of the Poynting vector. The four transverse field components derived from $u$ are:

$$E_{\theta\_u} = -\frac{i \cdot k}{r_s \cdot r} \cos\varphi \sum_{n=1}^{\infty} \frac{(2n+1)}{n(n+1)} [a_n \zeta_n'(k_0 r_s) - \varphi_n'(k_0 r_s)] \zeta_n'(k_0 r)[\cos\theta P_n'(\cos\theta) - \sin^2\theta P_n''(\cos\theta)],$$

(S6a)

$$E_{\varphi\_u} = \frac{i \cdot k}{r_s \cdot r} \sin\varphi \sum_{n=1}^{\infty} \frac{(2n+1)}{n(n+1)} [a_n \zeta_n'(k_0 r_s) - \varphi_n'(k_0 r_s)] \zeta_n'(k_0 r) P_n'(\cos\theta),$$ (S6b)

$$H_{\theta\_u} = \frac{k}{r_s \cdot r} \sin\varphi \sum_{n=1}^{\infty} \frac{(2n+1)}{n(n+1)} [a_n \zeta_n'(k_0 r_s) - \varphi_n'(k_0 r_s)] \zeta_n(k_0 r) P_n'(\cos\theta),$$ (S6c)

$$H_{\varphi\_u} = \frac{k}{r_s \cdot r} \cos\varphi \sum_{n=1}^{\infty} \frac{(2n+1)}{n(n+1)} [a_n \zeta_n'(k_0 r_s) - \varphi_n'(k_0 r_s)] \zeta_n(k_0 r)[\cos\theta P_n'(\cos\theta) - \sin^2\theta P_n''(\cos\theta)],$$

(S6d)

and those from $v$ are:

$$E_{\theta\_v} = \frac{i \cdot k}{r_s \cdot r} \cos\varphi \sum_{n=1}^{\infty} \frac{(2n+1)}{n(n+1)} [\varphi_n(k_0 r_s) - b_n \zeta_n(k_0 r_s)] \zeta_n(k_0 r) P_n'(\cos\theta),$$ (S7a)

$$E_{\varphi\_v} = -\frac{i \cdot k}{r_s \cdot r} \sin\varphi \sum_{n=1}^{\infty} \frac{(2n+1)}{n(n+1)} [\varphi_n(k_0 r_s) - b_n \zeta_n(k_0 r_s)] \zeta_n(k_0 r)[\cos\theta P_n'(\cos\theta) - \sin^2\theta P_n''(\cos\theta)],$$

(S7b)

$$H_{\theta\_v} = \frac{k}{r_s \cdot r} \sin\varphi \sum_{n=1}^{\infty} \frac{(2n+1)}{n(n+1)} [\varphi_n(k_0 r_s) - b_n \zeta_n(k_0 r_s)] \zeta_n'(k_0 r)[\cos\theta P_n'(\cos\theta) - \sin^2\theta P_n''(\cos\theta)],$$

(S7c)

$$H_{\varphi\_v} = \frac{k}{r_s \cdot r} \cos\varphi \sum_{n=1}^{\infty} \frac{(2n+1)}{n(n+1)} [\varphi_n(k_0 r_s) - b_n \zeta_n(k_0 r_s)] \zeta_n'(k_0 r) P_n'(\cos\theta).$$ (S7d)

The total transverse fields are thus

$$E_\theta = E_{\theta\_u} + E_{\theta\_v},$$ (S8a)

$$E_\varphi = E_{\varphi\_u} + E_{\varphi\_v},$$ (S8b)

and

$$H_\theta = H_{\theta\_u} + H_{\theta\_v},$$ (S9a)

$$H_\varphi = H_{\varphi\_u} + H_{\varphi\_v}.$$ (S9b)

Knowing the coefficients in Eqs. S4 and S5 leads to the outgoing Poynting vector

$$P_r = \frac{1}{2}\text{Re}(\mathbf{E} \times \mathbf{H}^*) \cdot \hat{r} = \frac{1}{2}\text{Re}(E_\theta H_\varphi^* - E_\varphi H_\theta^*). \tag{S10}$$

Although the rigorous solution contains an infinite number of spherical wave functions, in practice, one needs to truncate the series at a finite order. For the sphere sizes and wavelength range we are interested in, fortunately, a few low-order modes (N ≤ 4) can already produce good approximations of the radiation patterns simulated in CST (Studio Suite 2019), as shown in Fig. S4. Nevertheless, unless otherwise specified, we took $N = 15$ when plotting the patterns and calculating the F/B ratios. Finally, for the sake of comparison, the radiation patterns are plotted in a scale normalized to the maximum of $P_r$.

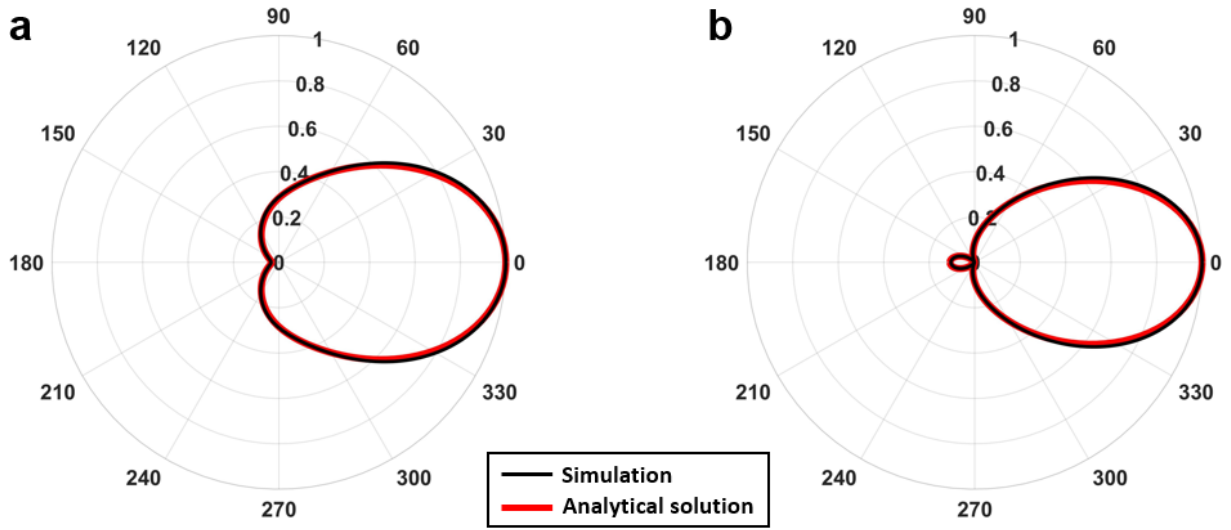

**Figure S4. Comparison of radiation patterns from full-wave simulations and from the analytical solution. a.** WS$_2$ exciton emission at 635 nm wavelength modulated by a 320 nm a-SiNS:H. **b.** MoS$_2$ exciton emission at 680 nm wavelength modulated by a 385 nm a-SiNS:H. Analytical solutions are truncated at $N = 4$ to validate the phasor diagrams in Figure 4 in the main text. Other patterns throughout this manuscript contain higher order modes with $N$ up to 15, contributing small correction terms to the rigorous solutions. No substrate included in both numerical and analytical approaches here.

## IV. Reciprocity theorem and simulations

The reciprocity theorem in electromagnetics relates "the field at one source due to a second source" to "the field at the second source due to the first source"[14,15], and it holds as long as the medium is linear, static and not biased by a quantity odd under time-reversal. The application of reciprocity theorem in antenna theory relates the transmitting and receiving properties of a radiating system. Therefore, although the excitation of excitons in TMDs and the light emission are two different processes, their directivity modulations share the F/B ratio if the wavelengths are the same. With this in mind, the modified Mie theory for dipole excitation (Section III) can instruct not only the modulation of emission but also the modulation of excitation in the same sphere-dipole system.

Similarly, in simulations, the F/B ratio can be determined by two means as presented in Fig. S5: (a) in the scenario of excitation, the incident plane waves launched from the top (forward excitation) and bottom (backward excitation) generate local fields recorded by a probe at the position of the monolayer TMD. The square of their ratio gives the F/B ratio. Note that only the tangential field components are taken for the calculation due to their alignment with the dipole orientation. (b) Alternatively, in the scenario of emission, the F/B ratio is obtainable from the radiation pattern by dividing the radiated power toward the top (forward emission) by that toward the bottom (backward emission).

Schematics of the models for simulating F/B based on the two scenarios are presented in Fig. S5. They are demonstrated totally equivalent, giving us identical F/B ratio mapping as a function of a-SiNS:H diameter and wavelength (right panel of Fig. S6 is based on the model in Fig. S5a, while Fig. 1c in the main text is based on the model in Fig. S5b).

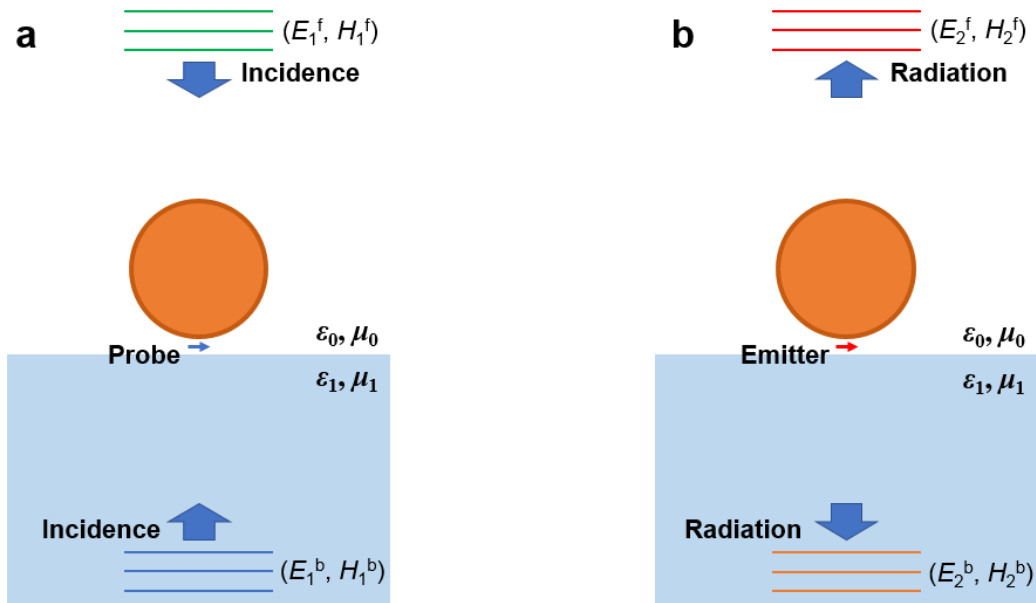

**Figure S5. Numerical simulation models for directivity modulation of both excitation and emission via an a-SiNS:H. a** modelling plane-wave-excited near field intensity at the position of dipole emitter. **b** modelling dipole-emitter-excited far-field radiation intensity**.** At a fixed wavelength, **a** and **b** are reciprocal to each other, and thus give the same F/B ratios.

## V. Mapping of F/B ratios by analytical model

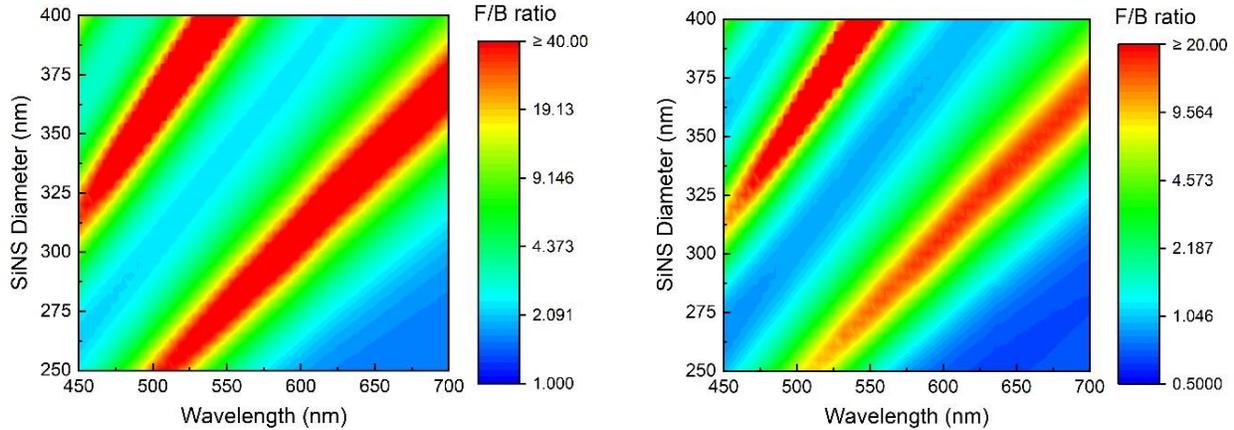

**Figure S6. Analytically (left) and numerically (right) calculated F/B ratio mapping** as a function of a-SiNS:H diameter and wavelength. The numerical one includes the glass substrate whereas the analytical one does not. The right panel is identical to Fig. 1c in the main text. The upper limits of the colormaps are respectively set to be 40 and 20 for better visibility. Higher values appear saturated in the maps.

In addition to the map of F/B intensity ratios extracted from the full-wave simulations for realistic structures with a substrate, we also did the mapping using our analytical model. As shown in Fig. S6, in the absence of the substrate, the directivity modulation by the a-SiNS:H size and wavelength follows a similar trend as the numerical results suggest.

## VI. Influence of the glass substrate on radiation patterns

To account for the substrate effect, we conduct full-wave simulations with the presence of a semi-infinite substrate of glass. The comparisons are shown in Fig. S7. Except two lobes pointing into the substrate, which are associated with the dipole's evanescent fields[16], the shape of the patterns are largely unchanged. However, when a glass substrate is there, the radiation will be guided towards high-index medium (substrate). Therefore, the backward component is increased, and the forward component is decreased, which agree with the mappings in Fig. S6. It should be noted that, no matter with or without a substrate, the reciprocity theorem always holds as we discussed in Section IV.

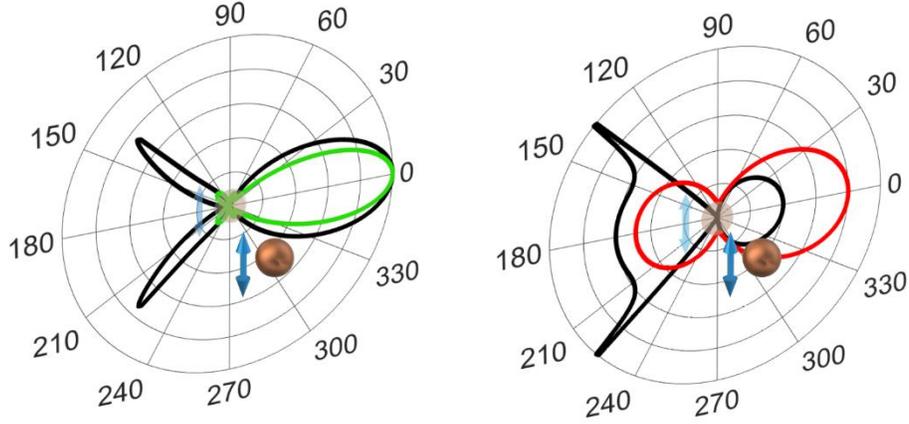

**Figure S7. Influence of the glass substrate on radiation patterns.** Examples of numerically simulated radiation patterns are illustrated in green (390 nm a-SiNS:H and 532 nm wavelength) and red (250 nm a-SiNS:H and 635 nm wavelength) according to Fig. 1a, b, respectively. The numerically simulated radiation patterns taking the glass substrate into account are also illustrated as black curves for comparison. The substrate is positioned at backward (180 degree) direction, adjacent to the dipole.

## VII. Low power approximation to effectively separate the modulation of excitation and emission

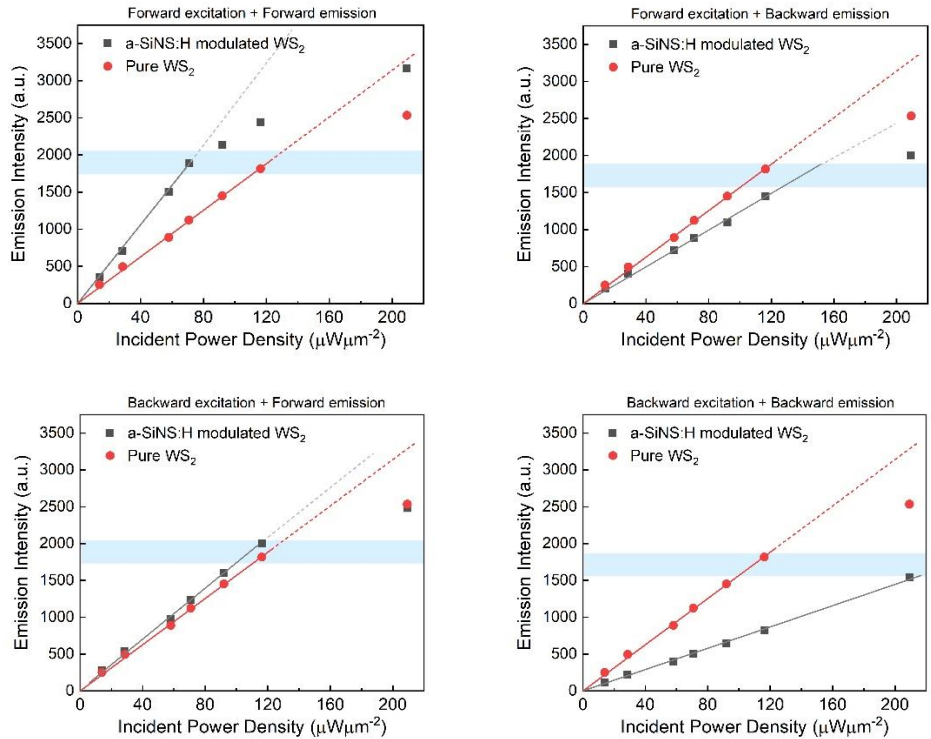

**Figure S8. Incident power density dependence of the emission intensity** for both pure $WS_2$ and a-SiNS:H modulated $WS_2$. Four possible combinations of forward/backward modulation of excitation and emission are listed based on 532 nm excitation and a 300 nm a-SiNS:H.

Choosing 532 nm laser excited monolayer WS$_2$ with a 300 nm a-SiNS:H modulator as an example, we experimentally prove that the modulated emission is always linearly proportional to the pure emission of bare WS$_2$ when the incident power density is lower than 70 µW/µm$^2$ (Fig. S8). In other words, a constant $\alpha_{F(B)}^{Ex(Em)}$ (See Eq. 3 in the main text) can be used to approximate this modulation function no matter for excitation or for emission processes. For the data in Figs. 2 and 3, an incident power density of 25.5 µW/µm$^2$ is applied for 532 nm laser. The saturation of emission signal at higher excitation powers can be attributed to exciton-exciton Auger processes[17,18].

**VIII. Measurements and data analysis for F/B ratios**

Our modified inverted microscope system includes two input light paths (from top and bottom) to focus the excitation laser on a-SiNS:H-TMD hybrids and one output path (toward bottom objective) to collect the emission signal, as shown in Fig. S9a (Fig. 2b). We use a 50X objective (N.A. = 0.60) with a working distance (WD) of 11.0 mm from top, and a 40X objective (N.A. = 0.65) with a WD of 0.66 mm from bottom. First, we achieve the best focus on the sample plane when laser is from bottom (40X), do the measurements and record its power and beam size. Then we let the laser come from top (50X) and make the power and beam size fit those from bottom by adjusting the focus plane and laser output power. Here, the 50X objective with a longer WD is important because it gives us freedom to adjust the focus plane and fit the focused beam size generated by the 40X objective from bottom. The beam size at the focal plane is kept at ~2 µm in diameter (Fig. S9a), no matter it is incident from top or bottom.

The comparison of the forward and backward modulated emission is carried out when the modulation of excitation is along a fixed direction, and vice versa for studying the directivity of excitation. For example, as shown in Fig. S9a, with sample facing down, we always collect the forward modulated emission signals from the bottom, and the value of F/B ratio ($R_{F/B}$) for excitation can thus be studied as

$$R_{F/B}|_{Ex} = \frac{\alpha_F^{Ex}}{\alpha_B^{Ex}} = \frac{(\alpha_F^{Em} \cdot \gamma \cdot \alpha_F^{Ex} \cdot P_1)/(\gamma \cdot P_1)}{(\alpha_F^{Em} \cdot \gamma \cdot \alpha_B^{Ex} \cdot P_2)/(\gamma \cdot P_2)} = \frac{Sig_F/Sig_F^0}{Sig_B/Sig_B^0}. \quad (S11)$$

$Sig_{F(B)}$ and $Sig_{F(B)}^0$ are the WS$_2$ emission signals to be collected with and without a-SiNS:H modulation, while the subscripts refer to (F) forward excitation and (B) backward excitation configurations (Fig. S9, right and left), respectively. $\gamma$ is a constant depending on the material and excitation wavelength, which describes the energy transfer from excitation to emission. However, it will not affect the F/B ratios at different wavelengths. In practice, despite the independence on incident power densities $P_1$ and $P_2$ under low power approximation (Section VII), we still maintain them at similar values for the sake of data accuracy.

As illustrated in Fig. S9b, the size of laser beam is larger than that of a-SiNS:Hs. Therefore, some excitons in the part of monolayer $WS_2$ not covered by a-SiNS:Hs are excited to emit light without coupling with a-SiNS:Hs. We regard them as the background emission signal. Because of their existence, the exact directivity should be much more significant if the monolayer $WS_2$ not covered by a-SiNS:Hs is etched[19] or a smaller N.A. is applied in signal collection. In other words, both forward and backward enhancement is less evaluated in our collected data.

Meanwhile, the existence of two substrate-introduced lobes on the backward side of far-field radiation pattern (Fig. S7, black curves) originally has few influences on the exact F/B ratio. However, as illustrated in Fig. S9c, they will be unwantedly collected by a large N.A (orange shaded area), as compared to the ideal case of a N.A. equals to zero (blue shaded area). Consequently, due to the diffraction limit of the detection as well, the forward enhancement is further less evaluated, whereas the backward enhancement is somehow over evaluated in our collected data.

In total, we would see a relatively good match on the value of F/B ratio ($R_{F/B}$) when it is smaller than 1 (backward enhancement), and a bad match for $R_{F/B} > 1$ case (forward enhancement, the larger the worse).

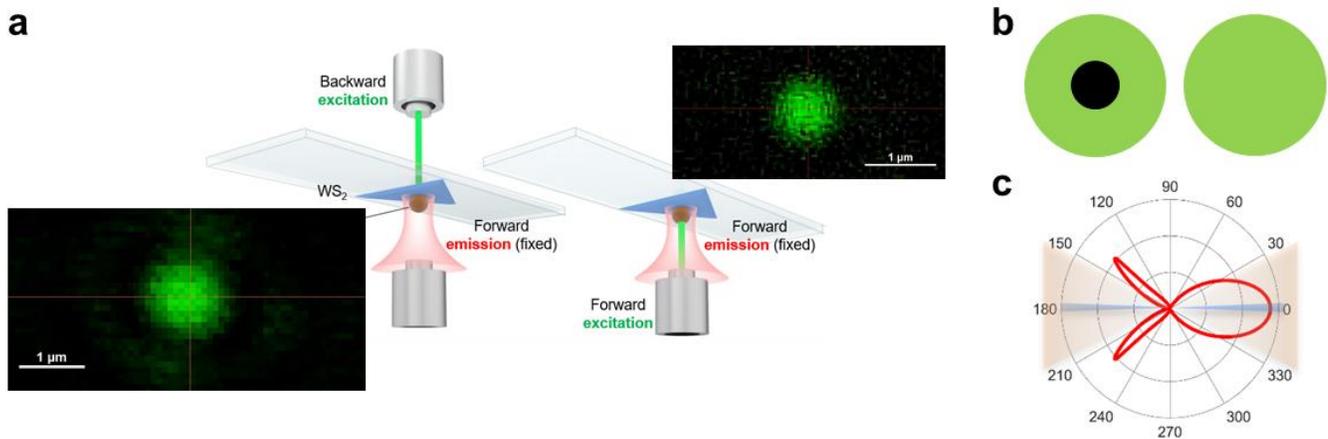

**Figure S9. Data acquisition and analysis for F/B ratios. a.** Experimental setups with the images of 532 nm excitation laser beam at the focal plane, when laser is incident from top (left) and bottom (right). **b.** Schematics showing the relative size of a-SiNS:Hs (black circle) and the laser beam (green circle). **c.** Typical far-field radiation pattern (the same as the black curve in the right panel of Fig. S7) when substrate effect is considered, showing the influence of N.A. on the collected signals by the objective. Blue shaded area: the ideal case when N.A. equals to zero, the same as how we define F/B ratio in simulations. Orange shaded area: an N.A. of 0.65 used in the experiments.

# IX. Converting the simulated F/B ratios into measured (predicted) values

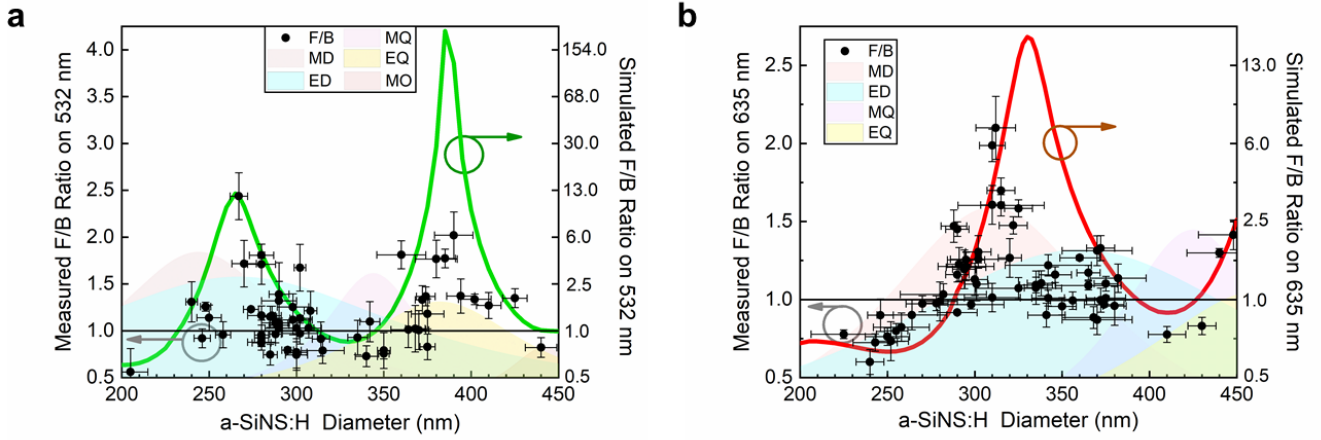

**Figure S10. Measured and simulated F/B ratios** as a function of a-SiNS:H diameter for both **a** 532 nm excitation and **b** 635 nm WS$_2$ emission. Full ranges of y-axes are displayed. The scales of simulated and measured F/B ratios follow the same relationship according to Eq. S12. The contributions of magnetic dipole (MD), electric dipole (ED), magnetic quadrupole (MQ), electric quadrupole (EQ), and magnetic octupole (MO) vary with the modulator size as well. They are displayed in the shaded backgrounds.

As illustrated in Fig. 2c (S10a) and 2d (S10b), the simulated F/B ratio is drawn in a logarithmic $y$ axis, showing the well-fitted tendency clearly together with experimental data. According to the discussions in Section VIII, we would see a relatively good match on the value of F/B ratio ($R_{F/B}$) when it is smaller than 1, and a bad match for $R_{F/B} > 1$ case (the larger the worse). Moreover, such a systematic mismatch is wavelength independent. Therefore, if we set the experimental F/B values as a standard, an empirical formula could always be found to transfer the simulated values into predicted values which involves the minimum errors in both Fig. S10a and b. Here we use a logarithmic function as an approximation to simplify this transfer formula as follow,

$$R_{pred} = \frac{5 \log_{60}(2 R_{sim}) + 1}{2}, \qquad (S12)$$

where $R_{sim}$ and $R_{pred}$ are the simulated and predicted (measured) F/B ratios, and the parameters are optimized based on the minimum errors in both Fig. S10a and b. The significance of introducing Eq. S12 is to give a reasonable instruction on a-SiNS:H's modulation of any dipole emitters based on our numerical and analytical predictions (Fig. S6).

We also notice the highest peak at ~390 nm a-SiNS:H in Fig. S10b, with a larger difference between the measured and simulated F/B ratios. Although it is experimentally demonstrated with 532 nm excitation, based on reciprocity theorem, we find that the observed "larger difference at larger a-SiNS:H size" is due to the existence of multiple dipole sources in practice. As a test, we set five dipoles, with the same orientation and phase, placed around the a-SiNS:H instead of just one in the simulation model (Fig. S11a). Compared to the one-dipole simulation result, we get a reduced $R_{F/B}$ with five dipole sources as shown in Fig. S11b. However, the reduction of $R_{F/B}$ is more obvious for large spheres (390 nm) than that for small spheres (260 nm). Moreover, taking the finite N.A. in experiments into account, larger spheres also show more lobes because high order modes are excited and come into play, which may further reduce the measured $R_{F/B}$ when it is larger than 1 as we discussed in Section VIII.

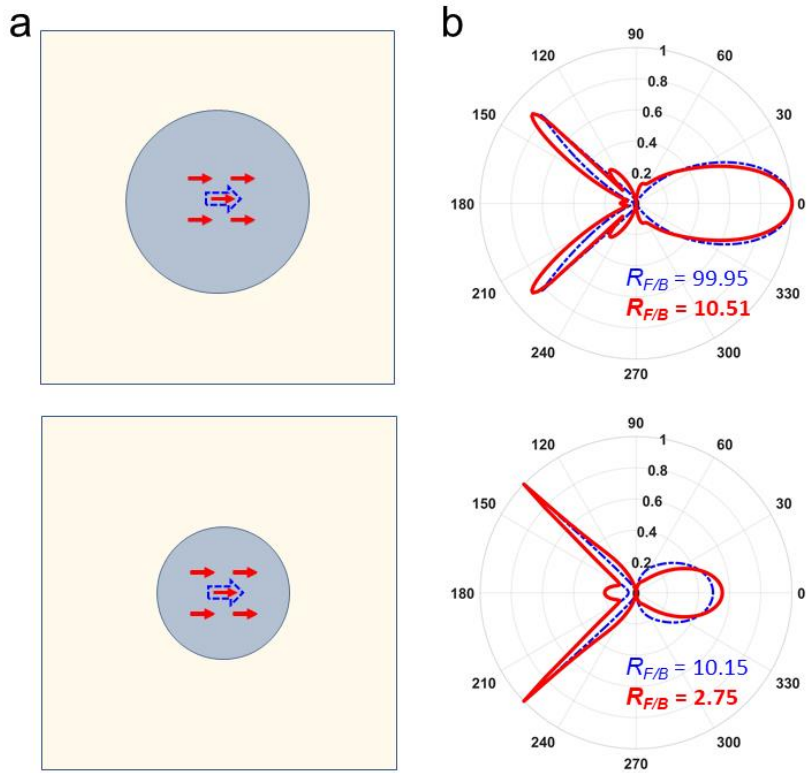

**Figure S11. Comparison of simulation results based on one dipole source and five dipole sources. a.** Schematics from top view showing 1 and 5 dipoles coupled with 390 nm (top) and 260 nm (bottom) a-SiNS:Hs. The extra 4 dipoles are translated from the sphere-substrate gap by 50 nm along the *x*- and *y*-axis. **b.** The corresponding comparisons of radiation patterns and F/B ratios when 1 dipole (blue dashed curves) and 5 dipoles (red curves) are considered, respectively.

## X. Total modulation of both excitation and emission

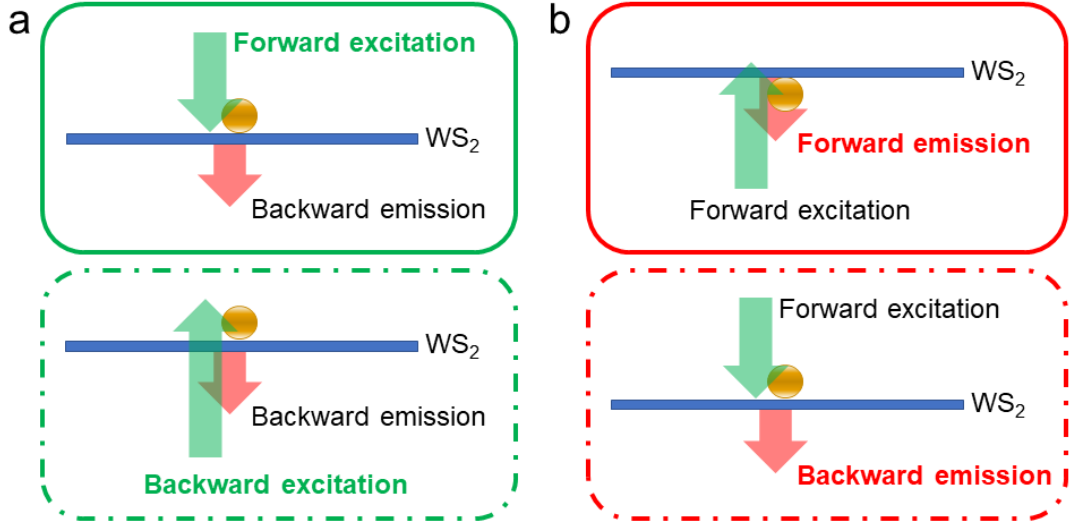

**Figure S12. Schematics of the two-step modulation** according to **a.** the solid and dashed green curves in Fig. 3b, and **b.** those red curves in Fig. 3d.

## XI. Reciprocity theorem in phasor diagrams

In the analysis where the substrate is absent, the reciprocity can be manifested by not only the F/B ratio (Section IV) but also the phasors. For an incoming plane wave from the forward (backward) direction, the normalized local field at the TMD from each multipole is identical to its dipole-emitted counterpart in the far-field region in the forward (backward) direction. (See the schematics in Fig. 1a, b in the main text.)

|  | plane wave excitation | dipole emission |
|---|---|---|
| Incident field | 1 | 1 |
| $a_1$ | 0.2244+0.1803i | 0.2244+0.1803i |
| $b_1$ | −0.4376−0.3473i | −0.4376−0.3473i |
| $a_2$ | −0.2148−0.6805i | −0.2148−0.6805i |
| $b_2$ | −0.0621+0.8624i | −0.0621+0.8624i |
| $a_3$ | −0.2389+0.2760i | −0.2389+0.2760i |
| $b_3$ | −0.1156−0.2721i | −0.1156−0.2721i |

**Table S1. Comparison of phasors** for the excitation process when the Mie resonances are triggered by plane-wave incidence and the emission process when the Mie resonances are triggered by a dipole source.

Since these two cases simply lead to the same phasor diagram, we show their equivalence by comparing the numerically calculated fields from individual Mie resonances under conditions reciprocal to each other. For the excitation process, the fields are evaluated at the location of the dipole and normalized to the backward incident plane wave at that point. For the emission process, the fields are evaluated at a distance of 10 μm from the a-SiNS:H in the backward direction and normalized to the dipole-emitted field at the same position. Table S1 shows the phasors of the normalized electric fields from the first 3 orders of $a_n$ and $b_n$ for 460 nm wavelength and a 320 nm a-SiNS:H, as an example (Fig. 4a in the main text).

## XII. Supplementary phasor diagrams

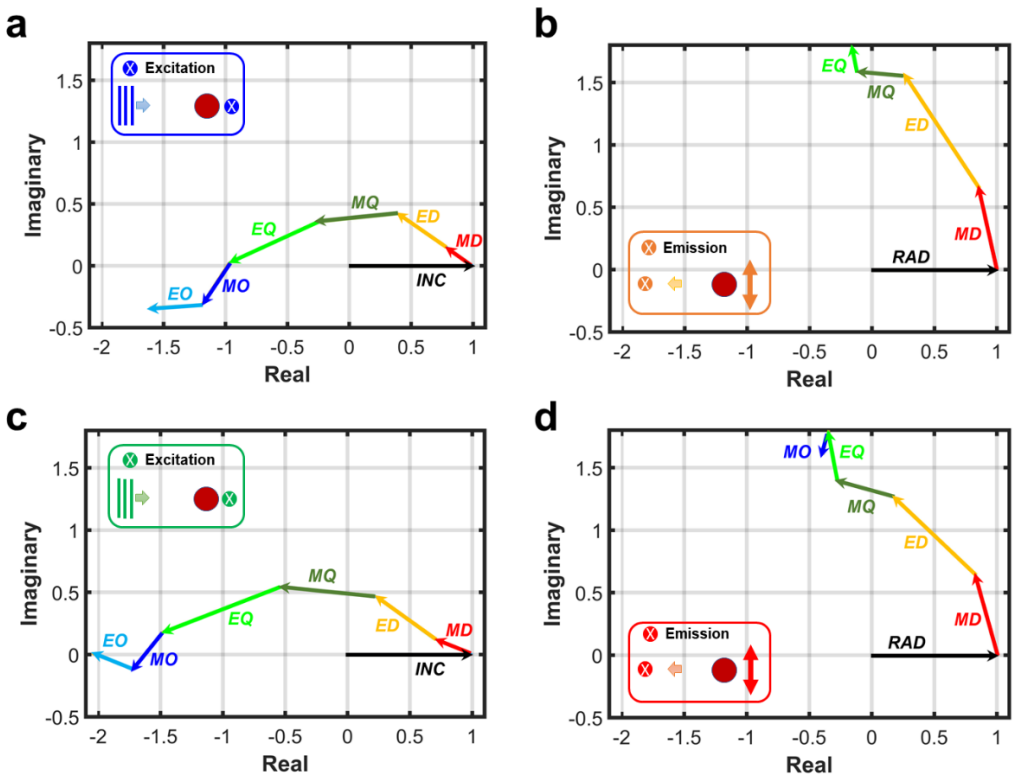

**Figure S13. Phasor diagrams showing the forward modulation** of both **a** 446 nm excitation and **b** 635 nm WS$_2$ emission via a 320 nm a-SiNS:H. The cross points in the insets denote the positions where the phasor is extracted. The incident field (INC), radiation field (RAD), resonant magnetic dipole (MD), electric dipole (ED), magnetic quadrupole (MQ), electric quadrupole (EQ), magnetic octupole (MO), and electric octupole (EO) are labelled for each vector. All of them are normalized based on INC/RAD. **c, d.** The same as **a, b**, but for 532 nm excitation and 680 nm MoS$_2$ emission via a 385 nm a-SiNS:H.

As the supplementary phasor diagrams to those in Fig. 4 in the main text, Fig. S13a, b are corresponded to Fig. 4a, b, while Fig. S13c, d are related to Fig. 4, d, e. They illustrate the forward components under a total modulation design for highly directional forward emission with maximum forward excitation efficiency.

## XIII. Emission control on different emitters and excitation wavelengths

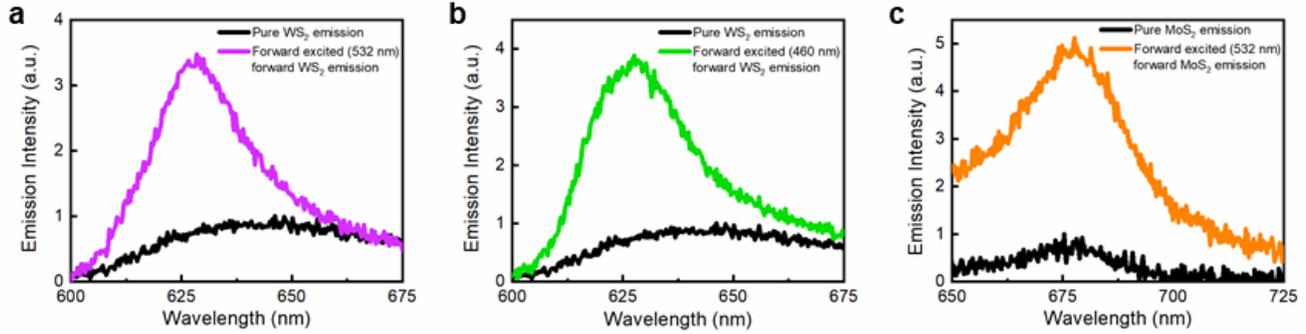

**Figure S14. Emission spectra of the cases when the largest total modulation (forward excitation + forward emission) is realized for different emitters and excitation wavelengths**. **a** A 390 nm a-SiNS:H modulator is chosen for 532 nm excited monolayer $WS_2$. **b** A 320 nm a-SiNS:H modulator is chosen for 446 nm excited monolayer $WS_2$ (the same as Fig. 4a, b). **c** A 385 nm a-SiNS:H modulator is chosen for 532 nm excited monolayer $MoS_2$ (the same as Fig. 4d, e). The excitation powers are **b** 1.02 $\mu W\mu m^{-2}$ for 446 nm laser and **a, c** 3.20 $\mu W\mu m^{-2}$ for 532 nm laser, respectively.

Different from the incident power density of 25.5 $\mu W\mu m^{-2}$ for 532 nm laser used in Fig. 3b, d, the power densities here are 1.02 $\mu W\mu m^{-2}$ for 446 nm laser and 3.20 $\mu W\mu m^{-2}$ for 532 nm laser in Fig. S14 and Fig. 4c, d. Therefore, the signal to noise ratio is lower and the peaks are broader. The broader shoulder of the monolayer $WS_2$ emission at longer wavelength (Fig. S14a, b) comes from the transfer of trion emission to localized states at ultra-low excitation densities[20]. Similar broader shoulder at longer wavelength can also be observed in the monolayer $MoS_2$ case (Fig. S14c) while a tail of B exciton emission at shorter wavelength is more obvious.

Since we study the directivity modulation (F/B ratio) of the emission at 635 nm (680 nm) wavelength via the a-SiNS:H specifically, no matter what's included in the total emission intensity, the intensity ratios at 635 nm (680 nm) are directly subtracted from Fig. S14a, b (Fig. S14c).

Moreover, as a supplementary to the 25.5 $\mu W\mu m^{-2}$ we used for the data in Fig. 2 and 3, the good matching between the measured and predicted values under such a low power density as shown in Fig. 4c, f also proves the compatibility of our model on different incident power densities according to Fig. S8.

A video according to Fig. S14a (532 nm excited 635 nm WS$_2$ emission modulated by a 390 m a-SiNS:H) is provided. It follows a total modulation of forward excitation and forward emission as shown in the schematic below:

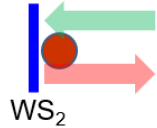

A 100X objective (N.A. = 0.90, WD = 1.0 mm) is used instead of the 40X one (N.A. = 0.65, WD = 0.66 mm) that we used for all the other data in this manuscript. The purpose is to optimize the video quality. The incident power density is 12.5 µWµm$^{-2}$. In the video, we scanned the laser spot around the a-SiNS:H but always shined it on a monolayer WS$_2$ flake. The total modulation of both excitation (green) and emission (red) is very obvious: (i) More excitation light is redistributed towards the WS$_2$ flake via the a-SiNS:H and thus almost no reflected green laser can be seen; (ii) More emission light is redistributed towards the objective via the a-SiNS:H and thus we can collect the strong red (635 nm) signal.

## XIV. Ultra-compact and multiplexed integrated photonics

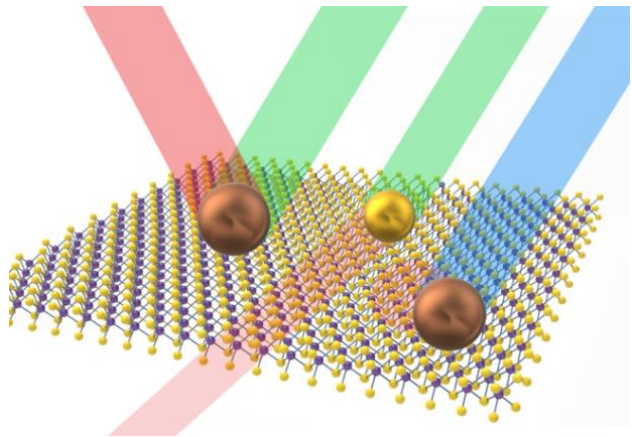

**Figure S15. Truly subwavelength a-SiNS:Hs for ultra-compact and multiplexed integrated photonics.**

Based on what we have demonstrated in Fig. 4 in the main text, with rational design, a-SiNS:H can be used to develop ultra-compact and multiplexed devices (e.g. designs for enhanced emission (response) signal along the excitation (trigger) light direction or opposite to it).

Multiple a-SiNS:Hs on a single monolayer TMD flake can function as multiplexed and integrated photonic devices. To be more specific, the neighboring a-SiNS:H modulators can handle different functions on one TMD flake and the same a-SiNS:H can also be multifunctional by tuning the trigger signal wavelength, as illustrated in Fig. S15. For example, the left brown sphere with a green incident signal can generate a highly directional red signal, "reflecting" back, while a smaller yellow sphere in the middle can work in a "transmitting" mode instead. Moreover, the same brown sphere but with a blue incident signal will instead functionalize as a beam dump, with almost no red signal generated.

Overall, the effective directivity modulation of emitters can enable the high-quality light signal modulation in integrated photonics, where spatial resolution and signal directivity are very important for its compactness. Two degrees of freedom, i.e., the a-SiNS:H size and incident signal wavelength, of the modulation give us enough flexibility to tailor the secondary emission signal with demanded functions.